\documentclass[aps,epsfig,twocolumn,showpacs,superscriptaddress,floatfix]{revtex4-1}
\usepackage{graphicx} 
\usepackage{epsfig} 
\usepackage{dcolumn}
\usepackage{color}
\usepackage{tikz}
\usepackage{bm}
\usepackage{amsmath}

\begin{document}

\title{Structure and decay of $^{21}$B resonances}
\author{R. \'Alvarez-Rodr\'{\i}guez} 
\email[e-mail: ]{raquel.alvarez@upm.es}
\affiliation{
  ETSAM, Universidad Polit\'ecnica de Madrid, 
  Avda. Juan Herrera 4, E-28040 Madrid, Spain}
\author{E. Garrido}
\affiliation{Instituto de Estructura de la Materia, IEM-CSIC, 
  Serrano 123, E-28006 Madrid, Spain} 
\date{\today}

\begin{abstract}
  The structure of the neutron-rich nucleus $^{21}$B is investigated
  within a three-body model consisting of a $^{19}$B core and two
  neutrons. The three-body wave functions are computed using the
  hyperspherical adiabatic expansion method combined with the
  complex-scaling technique. Four potential parametrizations
  compatible with the available experimental information on $^{20}$B
  are considered. The calculations predict a low-lying $\frac{3}{2}^-$
  state below the experimentally observed resonance at 2.47~MeV. An
  analysis of the dependence on the three-body interaction shows that
  this state cannot be shifted to the experimental resonance energy;
  instead, it remains either bound or becomes a near-threshold
  resonance with a rapidly increasing width. These results indicate
  that the observed resonance is more naturally interpreted as the
  second $\frac{3}{2}^-$ state, whereas a lower-lying $\frac{3}{2}^-$
  state, dominated by an $s_{1/2}$ configuration, has so far eluded
  experimental observation.  Momentum correlations are obtained for
  the experimentally observed resonance. In addition, the existence of
  other resonances is investigated.
  \end{abstract}

\maketitle

\section{Introduction}

Magic numbers are known to play a crucial role in nuclear
physics. They correspond to the number of neutrons or protons that
completely fill a nuclear shell, thereby providing additional
stability compared to neighboring nuclei. Although these numbers,
which lead to closed-shell configurations, are well established along
the valley of stability in the nuclear chart, it is also well known
that they can change significantly when approaching the proton or
neutron driplines \cite{bro01,bro22}. For example, a recent mass
measurement of the proton-dripline nucleus $^{22}$Si has revealed the
appearance of a new magic number, $Z$=14, in this region of the
nuclear chart \cite{xin25}.

In the neutron dripline, one of the most studied regions corresponds
to the neutron number $N$=16, where signatures of a new magic number
have been identified \cite{oza00,kan13}. In fact, evidence for the
double magicity, in both neutrons and protons, of $^{24}$O has been
reported \cite{jan09,hof09,kan09}. The immediately lighter isotone,
$^{23}$N, exhibits a similar $N$=16 gap, showing only a small
reduction compared to that in $^{24}$O [9]. A similar situation occurs
in $^{22}$C, whose Borromean halo structure has been attributed to the
persistence of the $N$=16 shell closure \cite{sou16,sun18,sin19}.

The next nucleus in the $N$=16 isotone chain is
$^{21}$B. Experimentally, the search for this nucleus is particularly
challenging due to its exceptional $A/Z$ ratio. Although an infinite
nuclear-matter model predicted $^{21}$B to be particle-stable
\cite{nay99}, an initial attempt to observe it, reported in
Ref.\cite{oza03}, found no evidence of its existence and instead
suggested that $^{21}$B is particle-unstable. More recently, both
$^{20}$B and $^{21}$B have been observed for the first time at RIKEN
\cite{leb18}. The particle-unstable $1^-$ and $2^-$ doublet was
identified in $^{20}$B together with a higher lying
resonance. Regarding $^{21}$B, it was found to undergo direct
two-neutron decay, and the only state compatible with the experimental
data appears to be a $\frac{3}{2}^-$ resonance at 2.47$\pm$0.19 MeV
above the two-neutron emission threshold, which is assumed to be its
ground state. No signatures of a bound state were found in this
experiment.

The purpose of this work is to investigate the structure of $^{21}$B,
paying particular attention to the possible existence of a low-lying
bound state still compatible with the experimentally known resonance
reported in Ref.\cite{leb18}. The fact that this resonance decays
through two-neutron emission suggests a three-body description of
$^{21}$B as a $^{19}$B core with two valence neutrons. The
corresponding three-body wave functions are computed using the
hyperspherical adiabatic expansion method \cite{nie01}, supplemented
with a complex-scaling transformation \cite{ho83,moi98} in order to
obtain the three-body resonances in a clean and reliable way.

The paper is organized as follows. Section II describes the
construction of the two-body interactions, focusing on the four
potentials designed to model the neutron-$^{19}$B subsystem. In
Section III the theoretical framework based on the hyperspherical
adiabatic expansion method, together with the complex scaling
technique, is briefly outlined. The structure of the $\frac{3}{2}^-$
states of $^{21}$B are discussed in Section IV. Section V presents the
energy distributions for the three-body decay. Possible additional
low-lying resonances are explored in Section VI.  Finally, the main
results and conclusions of the present work are summarized in Section
VII.

\section{The two-body potentials}
\label{sec2}

The crucial ingredients governing the properties of the three-body
$^{21}$B nucleus are the underlying two-body potentials. The
neutron–neutron interaction is well known, and a simplified version of
it will be employed. The main source of uncertainty arises from the
neutron–$^{19}$B potential, since it must be constructed from the
limited information currently available on the properties of
$^{20}$B. For this reason, several versions of the neutron–$^{19}$B
potential, all consistent with the existing data on $^{20}$B, will be
developed. The details of these potentials are provided below.

\subsection{Neutron-neutron potential}

The neutron-neutron interaction considered here is the Gaussian
potential, with central, spin-orbit, and spin-spin terms, given in
Ref.\cite{gar04}. That is,
\begin{eqnarray}
&&V_{nn}(r) =  37.05\:e^{-(r/1.31)^2}
-7.38\:e^{-(r/1.84)^2} \nonumber \\
&&-23.77\: e^{-(r/1.45)^2} \bm {\ell}
  \cdot \bm s + 7.16\:e^{-(r/2.43)^2} S_{12}\nonumber \\
&& +\left( 49.40\: e^{-r/1.31)^2} +
29.53\:e^{-(r/1.84)^2}\right) \bm {s_1}
  \cdot \bm {s_2},
\end{eqnarray}
where $\bm{\ell}$ is the relative orbital angular momentum between the
two neutrons, whose spins, $\bm{s}_1$ and $\bm{s}_2$, couple to the
total spin $\bm{s}$.

This is a simple potential whose parameters -strengths and ranges- are
adjusted to reproduce the experimental $s$- and $p$-wave
nucleon-nucleon scattering lengths and effective ranges. No Coulomb
interaction is required in this case. This simplified version of the
neutron-neutron interaction is sufficient in order to describe very
weakly bound systems, where the details of the inner part of the
potential play a minor role.

\subsection{Neutron-$^{19}$B potential}

The interaction between the $^{19}$B-core and the neutron is modeled
using an $\ell$-dependent potential that also includes central,
spin-orbit and spin-spin components. To be precise, the form of the
potential is given by:
\begin{equation}
  V^{(\ell)} (r) = V_c^{(\ell)} (r) + V_{so}^{(\ell)}(r) \bm {\ell}
  \cdot \bm s_n + V_{ss} \bm s_c \cdot \bm j_n\:,
  \label{n19b}
\end{equation}
where $\bm {\ell}$ is the relative orbital angular momentum between
the neutron and the core, $\bm s_n$ is the spin of the neutron, $\bm
s_c$ is the spin of the core, and $\bm j_n = \bm \ell + \bm s_n$ is
the total angular momentum of the neutron. This choice of the
spin-orbit and spin-spin operators is particularly appropriate when
$\bm{s}_c\neq 0$ (the spin and parity of the $^{19}$B-core is
$\frac{3}{2}^-$), because the quantum numbers associated to the
potential in Eq.(\ref{n19b}) are consistent with the ones in a
shell-model description of the nucleus, making simpler the
implementation of the Pauli principle \cite{gar03}.

\begin{table}[t!]
\begin{tabular}{c|c|c}
  $J^\pi$ & $E$ (MeV) & $\Gamma$ (MeV) \\
  \hline
  \hline
  $1^-,2^- $  & 1.56$\pm$0.15 & $< 0.5$ \\
  $1^-,2^- $  & 2.50$\pm$0.09 & 0.9$\pm$0.3  \\
  $(0^-/3^-) $  & 4.86 & $<0.5$  \\
  \hline
\end{tabular}
\caption{Experimental energies, above the two-body threshold, and
  widths (both in MeV) of the $^{20}$B states reported in
  Ref.\cite{leb18}.}
\label{tablepotener}
\end{table}

The available experimental information about $^{20}$B has been
reported in Ref.~\cite{leb18} and collected in
Table~\ref{tablepotener}. In this work two low-lying states were
identified with energies 1.56$\pm$0.15 MeV and 2.50$\pm$0.09 MeV above
the neutron-$^{19}$B threshold, and widths smaller than 0.5 MeV and
0.9$\pm$0.3 MeV, respectively. Although their precise angular momentum
and parity was not assigned, they were identified as the members of
the ($1^-, 2^-$) doublet.  Together with these two states, an
additional one was observed at the resonance energy 4.86$\pm$0.25 MeV
above the two-body threshold, and width smaller than 0.5 MeV. Based on
shell-model calculations, the authors suggest that this higher state
may correspond to angular momentum and parity either $0^-$ or $3^-$.

According to the standard shell-model, the last neutron in $^{20}$B
should sit either in the 2$s_{1/2}$ or the 1$d_{3/2}$ shell (the two
unoccupied levels in the $sd$-shell). This implies that only $\ell=0$
and $\ell= 2$ relative angular momenta are relevant for the
description of the low-lying states. Furthermore, since the ground
state has been identified to be one of the two states in the ($1^-$,
$2^-$) doublet, and the $^{19}$B core has spin and parity
$\frac{3}{2}^-$, it is then natural to consider that the two states in
the ground state doublet arise from the coupling of the
$\frac{3}{2}^-$ angular momentum of the core and the neutron in a
relative $s$-wave (therefore with the neutron in the 2$s_{1/2}$
shell). In the same way, the coupling of the $\frac{3}{2}^-$ angular
momentum of the core and the neutron in a relative $d$-wave (the
neutron in the $d_{3/2}$ shell) will be responsible for the additional
$^{20}$B excited states.

Taking this into account the $s$-wave and $d$-wave neutron-$^{19}$B
potentials are constructed as follows:

\paragraph*{s-wave potential:} As previously discussed, the $s$-wave potential is constructed assuming that
the experimentally known doublet $(1^-,2^-)$ in $^{20}$B arises from
the coupling between the $\frac{3}{2}^-$ ground state of the
$^{19}$B-core and the neutron in the $s_{1/2}$ shell. In this way,
from Eq.(\ref{n19b}) one can easily see that the two-body potentials
describing the $1^-$ and $2^-$ states in $^{20}$B take the form:
\begin{equation}
  V_{1^-}(r) = S_c e^{-r^2/b_c^2} -\frac{5}{4} S_{ss} e^{-r^2/b_{ss}^2}
  \label{pot1-}
\end{equation}
whereas for $s_x=2$ it is given by
\begin{equation}
  V_{2^-}(r) = S_c e^{-r^2/b_c^2} +\frac{3}{4} S_{ss} e^{-r^2/b_{ss}^2},
  \label{pot2-}
\end{equation}
where we have chosen a Gaussian shape for the central and spin-spin
potentials with strength and range ($S_c$, $b_c$), and ($S_{ss}$,
$b_{ss}$), respectively.

\begin{table}

\begin{tabular}{c|  c c c c c}
  Waves & & I & II & III & IV\\
  \hline
  \hline
  $s$&$S_c^{(\ell=0)}$& $-6.35375$ & $-6.73625$ & $-6.35375$& $-6.73625$ \\
  & $b_c^{(\ell=0)}$ & 2.0 &   2.0 & 2.0 & 2.0\\
  &$S_{ss}^{(\ell=0)}$& 0.765 & $-0.765$ &  0.765& $-0.765$\\
  & $b_{ss}^{(\ell=0)}$ & 2.0 &   2.0 & 2.0 & 2.0\\
  \hline
  $d$   & $S_c^{(\ell=2)}$ & 24.6  & 24.6 & 14.1 & 14.1\\
    
        & $b_c^{(\ell=2)}$ & 2.0  &  2.0 & 2.0 &2.0\\
    
        & $S_{so}^{(\ell=2)}$ &80.0  & 80.0 & 80.0 &80.0\\
 
        & $b_{so}^{(\ell=2)}$ &2.0  &   2.0 & 2.0 & 2.0\\
       
        & $S_{ss}^{(\ell=2)}$ &7.0  & 7.0 & $-7.0$ &$-7.0$\\
       
        & $b_{ss}^{(\ell=2)}$ &2.0  &   2.0 & 2.0 & 2.0\\
  \hline
\end{tabular}
  \caption{Potential parameters for the central, spin-orbit and
    spin-spin Gaussian potentials for the neutron-$^{19}$B interaction
    (see Eq.\ref{n19b}).  Four different sets are considered. The
    units for the strengths $S^{(\ell)}$ and ranges $b^{\ell}$ are MeV
    and fm, respectively. The superscript $(\ell)$ refers to the
    different partial waves considered in this work. The subscripts
    $c$, $so$ and $ss$ indicate central, spin-orbit and spin-spin
    potentials, respectively.}
\label{tablepotparam}
\end{table}

The strength and range parameters are now adjusted to reproduce the
experimental energies of the ($1^-$, $2^-$) doublet.  Since these
states arise from a relative $s$-wave between the two particles, they
are then understood as virtual states, analogous to the virtual state
present in $^{10}$Li \cite{gar02}, and whose energy is estimated as
\cite{bet49}:
\begin{equation}
    E=\frac{\hbar^2 \kappa^2}{2\mu} \mbox{ with } 
    \kappa=\frac{1-\sqrt{1-\frac{2r_\mathrm{eff}}{a}}}{r_\mathrm{eff}},
\end{equation}
where $a$ ($<0$) and $r_\mathrm{eff}$ are, respectively, the
scattering length and the effective range of the two-body considered
potential.

Two scenarios consistent with the experimental data are possible,
depending on which of the two states in the doublet is assumed to be
the ground state. The strength and range parameters obtained for the
$s$-wave interaction are given in the upper part of
Table~\ref{tablepotparam}, where the strength and range of the
$s$-wave potential below the labels I and III (II and IV) correspond
to the $^{20}$B ground state with spin and parity $1^-$ ($2^-$). It is
important to note here that the strength parameters have been chosen
shallow enough to avoid the binding of the valence neutron into the
Pauli forbidden $1s_{1/2}$-state. An alternative could be to allow the
neutron to be bound into the forbidden state, and subsequently remove
this state as done for instance in Ref.\cite{gar99} by means of the
phase equivalent potentials.

The remaining details about the two-body potentials I, II, III, and IV
given in Table~\ref{tablepotparam} are determined by the $d$-wave
two-body interaction as described below.

\paragraph*{d-wave potential:} For an $\ell=2$ relative wave,
the neutron can sit either in a $d_{5/2}$ or a $d_{3/2}$-state. Since
the $1d_{5/2}$ shell is fully occupied by the neutrons in the core,
the strength of the spin-orbit potential in Eq.(\ref{n19b}) is chosen
such that the $d_{5/2}$-wave is pushed up in energy out of the active
shells. In this way the Pauli principle is preserved and the valence
neutron is mainly allowed in a $d_{3/2}$-state.

When coupled to the spin of core, $s_c = \frac{3}{2}$, this
configuration gives rise to a set of high-lying states ($0^-$, $1^-$,
$2^-$, and $3^-$) in $^{20}$B. The lowest of these states can be
either the $0^-$ or the $3^-$-state depending on the repulsive or
attractive character of the spin-spin potential. In this way we can
tentatively assign the spin and parity of the third state in $^{20}$B
to be either $0^-$ or $3^-$.

Following Eq.(\ref{n19b}), and choosing again a Gaussian shape for the
potentials, we obtain the following expressions for the the $d$-wave
potential leading to the $0^-$ and the $3^-$ states:
\begin{equation}
  V_{0^-}(r) = S_c e^{-r^2/b_c^2} - \frac{3}{2}S_{so} e^{-r^2/b_{so}^2}
  -\frac{15}{4} S_{ss} e^{-r^2/b_{ss}^2},
\end{equation}
and  
\begin{equation}
  V_{3^-}(r) = S_c e^{-r^2/b_c^2} - \frac{3}{2}S_{so} e^{-r^2/b_{so}^2}
  +\frac{9}{4} S_{ss} e^{-r^2/b_{ss}^2}.
\end{equation}
The specific values of the strength and range parameters used for the
$d$-wave potential are given in the lower part of
Table~\ref{tablepotparam}.

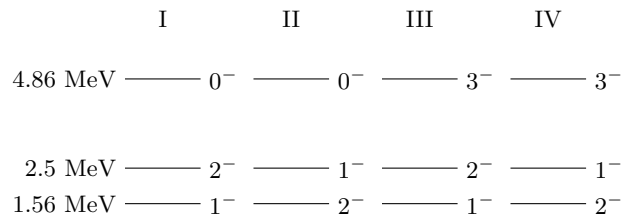
\begin{figure}
  \begin{tikzpicture}
\draw (0,1.56) node [left]{1.56 MeV} -- (1,1.56) node[right]{$1^-$};
\draw(0,2.03) node [left] {2.5 MeV} -- (1,2.03) node [right]{$2^-$};
\draw(0,3.21) node [left] {4.86 MeV}-- (1,3.21) node [right] {$0^-$};
\draw(0.5,4.) node {I};
\draw (1.7,1.56)  -- (2.7,1.56) node[right]{$2^-$};
\draw(1.7,2.03)  -- (2.7,2.03) node [right]{$1^-$};
\draw(1.7,3.21) -- (2.7,3.21) node [right] {$0^-$};
\draw(2.2,4.) node {II};
\draw (3.4,1.56)  -- (4.4,1.56) node[right]{$1^-$};
\draw(3.4,2.03) -- (4.4,2.03) node [right]{$2^-$};
\draw(3.4,3.21) -- (4.4,3.21) node [right] {$3^-$};
\draw(3.9,4.) node {III};
\draw (5.1,1.56)  -- (6.1,1.56) node[right]{$2^-$};
\draw(5.1,2.03)  -- (6.1,2.03) node [right]{$1^-$};
\draw(5.1,3.21) -- (6.1,3.21) node [right] {$3^-$};
\draw(5.6,4.) node {IV};

  \end{tikzpicture}
  \caption{Scheme of the energies (MeV) and spin-parities fitted by the four
    proposed neutron-$^{19}$B potentials.}
   \label{niveles}
\end{figure}

When combined with the $s$-wave potentials shown in the upper part, we
get four different potential sets, I, II, III, and IV, which
correspond to the four possible level schemes, consistent with the
limited experimental information available, shown in
Fig.~\ref{niveles}. Potentials I and III (II and IV) correspond to the
ground state in $^{20}$B to be the $1^-$ ($2^-$) state, and whereas
with potentials I and II the third state is a $0^-$-state, with
potentials III and IV it is a $3^-$-state.

\section{The three-body method}

The $^{21}$B nucleus is modeled as a three-body system composed of a
$^{19}$B core and two neutrons. In this work the three-body wave
functions are obtained after solving the Faddeev equations in
coordinate space by means of the hyperspherical adiabatic expansion
method described in detail in Ref.~\cite{nie01}.

\begin{figure}
  \begin{tikzpicture}
\shade[ball color = blue!40, opacity = 0.4] (0,0) circle (0.5cm);
\draw(0,0.7) node {core};
\shade[ball color = magenta!40, opacity = 0.4] (2,0) circle (0.15cm);
\draw(2,0.4) node {n};
\shade[ball color = magenta!40, opacity = 0.4] (1.3,-1) circle (0.15cm);
\draw(1.3,-1.4) node {n};
\draw(0,0)--(1.3,-1) node[midway,below]{$\bm x_3$};
\draw(2,0)--(0.5,-0.4) node [midway,above]{$\bm y_3$};
\draw(4,6) node {i=1};
\shade[ball color = blue!40, opacity = 0.4] (0,3) circle (0.5cm);
\draw(0,3.7) node {core};
\shade[ball color = magenta!40, opacity = 0.4] (2,3) circle (0.15cm);
\draw(2,3.4) node {n};
\shade[ball color = magenta!40, opacity = 0.4] (1.3,2) circle (0.15cm);
\draw(1.3,1.6) node {n};
\draw(0,3)--(2,3) node[midway, above] {$\bm{x_2}$};
\draw(1.3,2)--(0.6,3) node [midway,right]{$\bm y_2$};
\draw(4,3) node {i=2};
\shade[ball color = blue!40, opacity = 0.4] (0,6) circle (0.5cm);
\draw(0,6.7) node {core};
\shade[ball color = magenta!40, opacity = 0.4] (2,6) circle (0.15cm);
\draw(2,6.4) node {n};
\shade[ball color = magenta!40, opacity = 0.4] (1.3,5) circle (0.15cm);
\draw(1.3,4.6) node {n};
\draw(2,6)--(1.3,5) node[midway, right] {$\bm x_1$};
\draw(0,6)--(1.65,5.5) node[midway,above]{$\bm y_1$};
\draw(4,0) node {i=3};
  \end{tikzpicture}
  \caption{Scheme of the three different Jacobi sets of coordinates. The
  mass factors have been omitted for simplicity.}
   \label{jacobi}
\end{figure}
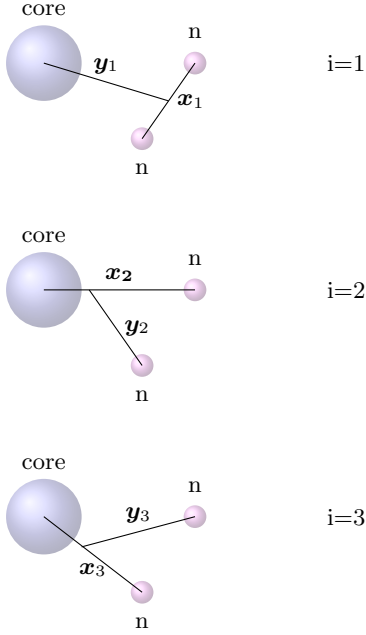

This method employs the hyperspherical coordinates defined after the
usual $\bm{x}$ and $\bm{y}$ Jacobi coordinates. Except for some mass
factors, see Ref.~\cite{nie01}, the $\bm{x}$-coordinate is the
relative distance between two of the particles, and $\bm{y}$ is the
relative distance between the third particle and center of mass of the
first two. The three possible choices for the Jacobi coordinates are
shown in Fig.~\ref{jacobi}, and they represent the Jacobi coordinates
used on each of the Faddeev components of the three-body wave
function. In the following we shall refer to the first Jacobi set as
the one with the $\bm{x}$-coordinate between the two neutrons (upper
part in Fig.~\ref{jacobi}), and to the second and third Jacobi sets as
the ones with the $\bm{x}$-coordinate between the core and one of the
neutrons (middle and lower parts in Fig.~\ref{jacobi}).

From the Jacobi coordinates one can construct the hyperspherical
coordinates, which contain a single radial coordinate, the
hyperradius, $\rho=(x^2+y^2)^{1/2}$, which is independent of the
choice made for the Jacobi coordinates, and five generalized
hyperangles, $\Omega_i \equiv
\{\alpha_i=\arctan{x_i/y_i},\Omega_{x_i},\Omega_{y_i} \}$, where
$i=1,2,3$ runs over the three possible definitions of the Jacobi
coordinates depicted in Fig.~\ref{jacobi}.

The total three-body wave function for a given angular momentum $J$
and projection $M$ is then expanded in terms of the basis set
$\{\Phi_n^{JM}(\rho,\Omega)\}$ as:
\begin{equation}
\Psi^{JM} = \frac{1}{\rho^{5/2}}\sum_n f_n^J(\rho) \Phi_{n}^{JM}(\rho,\Omega)\;,
\label{3bdwf}
\end{equation}
where the angular functions
\begin{equation}
\Phi_n^{JM}(\rho,\Omega)=
\sum_{i=1}^3 \phi_{n}^{JM(i)}(\rho,\Omega_i)
\label{angfun}
\end{equation}
are the eigenfunctions of the angular part of the Faddeev equations. 

The radial functions, $f_n^J(\rho)$, are obtained as the solutions of
the coupled set of differential equations
\begin{eqnarray}
\lefteqn{  \hspace*{-12mm}
\left[-\frac{d^2}{d\rho^2}+\frac{\lambda_n(\rho)+\frac{15}{4}}{\rho^2} -\frac{2m(E-V_\mathrm{3b}(\rho))}{\hbar^2} \right]f_n^J(\rho)= }
\nonumber \\ &&
=\sum_{n'} \left( 2 P_{nn'} (\rho) \frac{d}{d\rho}+Q_{nn'}(\rho) \right) f_{n'}^J(\rho),
\label{couprad}
\end{eqnarray}
where the eigenvalues of the angular part, $\lambda_n(\rho)$, enter as
effective potentials, and where the $P_{nn'}(\rho)$ and
$Q_{nn'}(\rho)$ functions, whose precise definition can be found in
\cite{nie01}, couple the different radial functions. In
Eq.(\ref{couprad}) $m$ is the normalization mass used to define de
Jacobi coordinates \cite{nie01}, $E$ is the three-body energy, and
$V_\mathrm{3b}(\rho)$ is the three-body potential, assumed to depend
on the hyperradius only, included to take care of those effects that
go beyond the bare particle-particle interactions.

As mentioned, the details of the method are given in
Ref.~\cite{nie01}, but let us simply specify here that the angular
eigenfunctions, $\Phi_n^{JM}(\rho,\Omega)$, are obtained after
expanding each of the three $\phi$ functions in Eq.(\ref{angfun}) as:
\begin{equation}
\phi_n^{JM}(\rho,\Omega)=\sum_{K,q} C_{n,q}^{JM}(\rho)\left[{\cal Y}_{\ell_x \ell_y}^{KL}(\Omega) \otimes \chi_{s_x s_y}^S \right]^{JM},
\label{angexp}
\end{equation}
where $q$ groups the quantum numbers $\{\ell_x, \ell_y, L,s_x,
S\}$. For each of the three Jacobi sets in Fig.~\ref{jacobi}, $\ell_x$
and $\ell_y$ are the orbital angular momenta associated to the
$\bm{x}$ and $\bm{y}$ coordinates, respectively, and they couple to
the total orbital angular momentum $L$. Also, the spins of the two
particles connected by the $\bm{x}$-coordinate couple to $s_x$, which
after coupling to the spin, $s_y$, of the third particle gives rise to
the total spin $S$. Finally, $L$ and $S$ couple to the total angular
momentum, $J$, of the three-body system. The hypermomentum quantum
number $K$ is defined as $K=2\nu+\ell_x+\ell_y$, with
$\nu=0,1,2,\cdots$, the functions ${\cal Y}_{\ell_x
  \ell_y}^{KL}(\Omega)$ are the usual hyperspherical harmonics, and
$\chi_{s_x s_y}^S$ represents the three-body spin part of the wave
function.

 The choice of the components $q\equiv\{\ell_x, \ell_y, L,s_x, S\}$
 and the maximum value of the hypermomentum $K$, $K_\mathrm{max}$,
 included in the expansion (\ref{angexp}) is a crucial ingredient of
 the calculation. It is necessary to make sure that the relevant
 components, $q$, are included, and also that $K_\mathrm{max}$ is
 large enough to ensure the convergence of the
 calculation. Additionally, in order to get convergence, a
 sufficiently high number of adiabatic terms must also be included in
 Eq.(\ref{3bdwf}). Typically, this is achieved after inclusion of a
 relatively small number of terms, usually no more than four or five.

The method described so far permits to obtain bound states simply
after imposing to the radial wave functions, $f^J_n(\rho)$, in
Eq.(\ref{3bdwf}), an asymptotic exponential fall off. However, due to
the non-integrability of the resonance wave functions, resonant states
can not be identified so easily. This problem is solved after
implementing a complex scaling transformation \cite{ho83,moi98} into
the hyperspherical adiabatic expansion method, as done for instance in
Ref.~\cite{fed03}. In this way, the energy and width of a resonance
are associated to the complex eigenvalues of an analytically continued
Hamiltonian operator.

More specifically, the transformed Hamiltonian results from the
rotation of the position vectors, $\bm{r}$, of the ordinary
Hamiltonian into the complex plane, i.e., all the radial coordinates,
$\bm{r}$, are replaced by $\bm{r}e^{i\theta}$ ($0<\theta<\pi/2$). As
shown in \cite{ho83,moi98}, resonances, understood as poles of the
$S$-matrix in the fourth quadrant of the energy plane, appear as
eigenfunctions of the complex rotated hamiltonian, $H_\theta (\bm r) =
H (\bm{r} e^{i\:\theta} )$, with complex eigenvalues. The important
point is that, provided the argument of the resonance is smaller than
$\theta$, the complex rotated resonance wave function falls off
exponentially, similarly to a bound state wave function. One can
therefore, after the complex scaling transformation, use for
resonances the same numerical techniques as for bound states. The
complex energy of the resonance, $E=E_R-i\frac{\Gamma_R}{2}$, permits
to extract both, the resonance energy, $E_R$, and the resonance width,
$\Gamma_R$. It is important to note that, after a complex scaling
transformation, the true bound states of the system are still found at
the correct energy.

\section{The $\frac{3}{2}^-$ states in $^{21}$B}

Following the procedure described in the previous section, and using
the two-body potentials given in Section~\ref{sec2}, we now proceed to
calculating the $\frac{3}{2}^-$ states in $^{21}$B.

The calculations are performed including in the expansion
(\ref{angexp}) all the components with $\ell_x,\ell_y \leq 3$ in the
first Jacobi set, and $\ell_x,\ell_y\leq 2$ in the second and third
Jacobi sets. In this expansion the maximum value of the hypermomentum,
$K_\mathrm{max}$, is always taken at least $K_\mathrm{max}$=50. For
the most relevant components it is increased up to
$K_\mathrm{max}$=250. Note that the components with a relative
$p$-wave between the neutron and the $^{19}$B-core are also included
in the calculation. Since there is no experimental information about
the existence of $p$-wave states in $^{20}$B (or they are in any case
expected to be high in energy) the results have been obtained putting
the $p$-wave neutron-$^{19}$B potential equal to zero.

The calculations are made after a complex scaling transformation with
rotation angle $\theta$=0.2~rad. The three-body spectrum is obtained
imposing a box boundary condition at a sufficiently large value of
$\rho$. In this way not only the bound states or resonances are
obtained, but also the continuum spectrum is computed, appearing as a
series of discrete states rotated in the fourth quadrant of the energy
plane by an angle $2\theta$ \cite{ho83,moi98}.

\begin{figure}
   \begin{center}
    \includegraphics[width = 8.5cm]{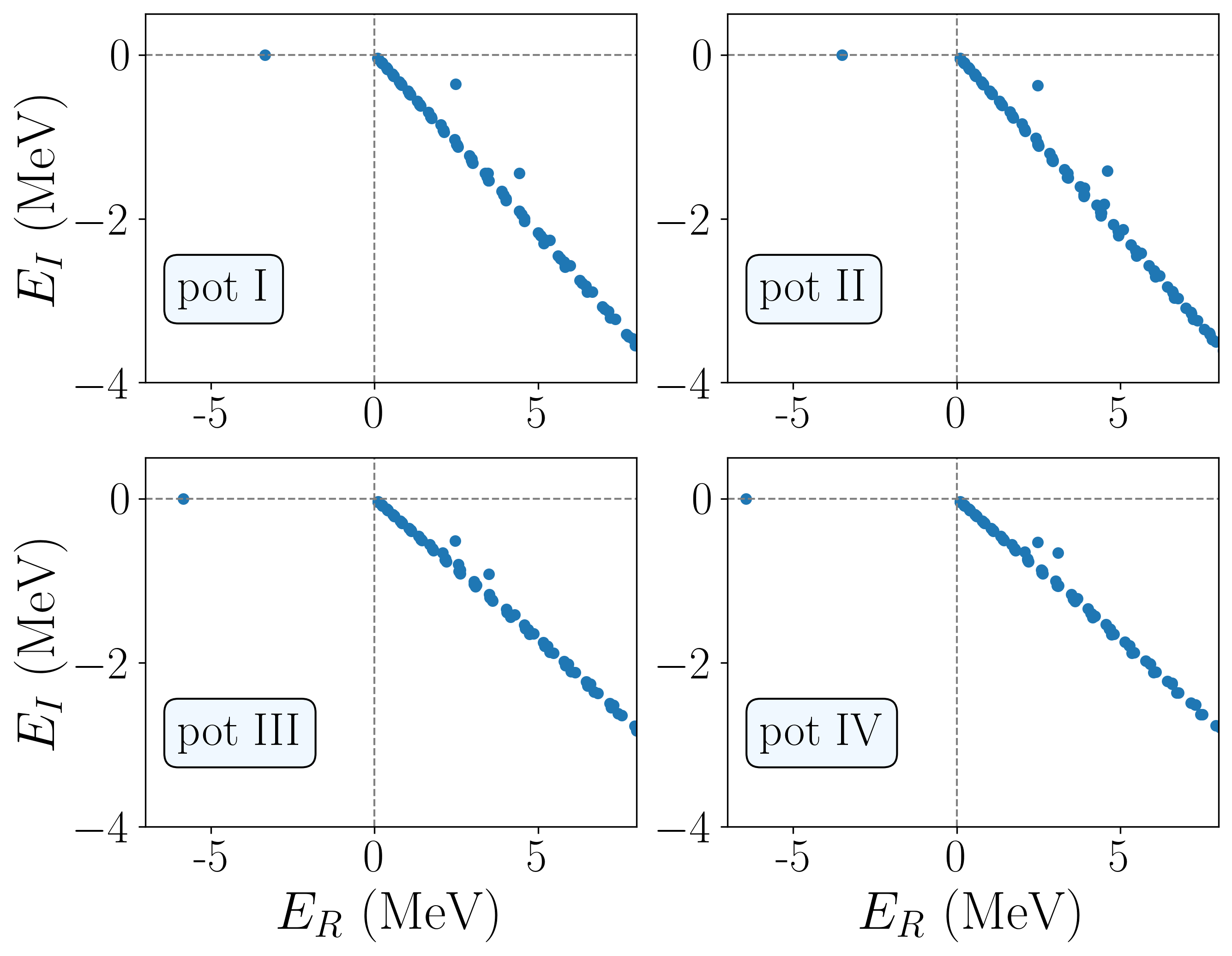}
   \end{center}
   \vspace*{-0.5cm}
   \caption{The real ($E_R$) and imaginary ($E_I$) parts of the energies
     for $\frac{3}{2}^-$ states in $^{21}$B obtained after complex rotation
     for the four (labeled as pot I-IV) potentials under
     consideration.  
}
   \label{spectr}
\end{figure}

The results of the calculations are shown in Fig.~\ref{spectr} for the
four potentials given in Table~\ref{tablepotparam}. In the
calculations a Gaussian three-body potential,
\begin{equation}
  V_\mathrm{3b}=S_\mathrm{3b}\exp (-\rho^2/b_\mathrm{3b}^2)
\label{3bpot}
\end{equation}
has been included in the radial equations (\ref{couprad}), such that
the lowest computed $\frac{3}{2}^-$ resonance appears, for each of the
four potentials, at the experimental value of 2.47 MeV
\cite{leb18}. To be precise, taking $b_\mathrm{3b}$=5 fm, the
strengths values $S_\mathrm{3b}$=$-11.9$, $-11.9$, $-16.6$, and
$-17.4$~MeV have being used for potentials I, II, III, and IV,
respectively.

In all the four panels shown in Fig.~\ref{spectr} we can see the
energy cut, starting at the origin, rotated by an angle $2\theta$,
that corresponds to the discretized continuum $\frac{3}{2}^-$
states. The dots out of this energy cut are either bound states, when
located in the negative energy axis, or three-body resonances, when
located in the fourth quadrant of the energy plane. As mentioned
before, the lowest of the resonances has been put at the experimental
value by means of the three-body potential in Eq.(\ref{3bpot}).

\begin{table}[t!]
\begin{tabular}{c| c c| c c| c c| c c}
  $J^\pi$ & $E_\text{I}$ & $\Gamma_\text{I}$ & $E_\text{II}$ &$\Gamma_\text{II}$&
  $E_\text{III}$&$\Gamma_\text{III}$ & $E_\text{IV}$ & $\Gamma_\text{IV}$ \\
  \hline
  \hline
  $\frac{3}{2}^-_1$  & $-3.35$ & 0    &$-3.50$ &0 &$-5.84$ &0 &$-6.44$  &0\\
  $\frac{3}{2}^-_2$  & 2.47  & 0.71 &2.47  &0.75 &2.47  & 1.03 &2.47  &1.06 \\
  $\frac{3}{2}^-_3$  & 4.42  & 2.89 &4.60 & 2.83  & 3.49  & 1.83  & 3.10& 1.31\\
  \hline
\end{tabular}
  \caption{Energies (MeV) and widths (MeV) of the three lowest
    $\frac{3}{2}^-$ states of $^{21}$B obtained with the four
    potentials under consideration, labeled as I-IV. The range of the
    three-body potential, Eq.(\ref{3bpot}), is taken to be
    $b_{3b}$=5~fm and the strengths are different for each of the four
    potentials under consideration, namely
    $S_\mathrm{3b}^\mathrm{I}=-11.9$, $S_\mathrm{3b}^\mathrm{II}=
    -11.9$, $S_\mathrm{3b}^\mathrm{III}=-16.6$ and
    $S_\mathrm{3b}^\mathrm{IV}=-17.4$~MeV). }
\label{tableen}
\end{table}

In Fig.~\ref{spectr} we can immediately see that, in all the four
cases, more than one $\frac{3}{2}^-$ resonance is found, and the
lowest of them can be located at the experimental value by means of
the appropriate three-body potential. The relevant thing is that,
together with the resonances, also in the four cases, a bound state is
also found.

The precise energies and widths of the computed $\frac{3}{2}^-$ states
with the four $n$-$^{19}$B potentials used in this work are given in
Table~\ref{tableen}. The width obtained for the second $\frac{3}{2}^-$
state is larger than the experimental estimate of $\Gamma < 0.6$ MeV
given in Ref.~\cite{leb18}, although for potentials I and II not
dramatically bigger. We also see that potentials III and IV (which
correspond to a high-lying $3^-$ state in $^{20}$B) produce a bound
state a couple of MeV more bound than potentials I and II (which
correspond to a high-lying $0^-$ state in $^{20}$B).

From the results shown in Fig.~\ref{spectr} and Table~\ref{tableen}
one could conclude that the ground state in $^{21}$B should actually
be a bound state, even if that state has so far not been observed
experimentally. However, in order to evaluate how strong this
conclusion is it is necessary to have a closer look into the
properties and behavior of the computed states.

\begin{table}
\begin{tabular}{cc|cccc}
$^2n$($\ell_x$, $j_x$) & $\ell_{j_y}$ &$W^\mathrm{I}$ &  $W^\mathrm{II}$  & $W^\mathrm{III}$  &  $W^\mathrm{IV}$\\ \hline \hline
$^2n$$(0, 0)$  & $s_{3/2}$  &  99.5  &  99.5   &  99.2  &   99.1  \\
  &   &   &    &   &   \\
$^{20}$B($\ell_{j_n}$, $j_x$) & $\ell_{j_y}$ 
&$W^\mathrm{I}$ &  $W^\mathrm{II}$  & $W^\mathrm{III}$  &  $W^\mathrm{IV}$\\ \hline \hline
$^{20}$B$(s_{1/2}, 1)$  & $s_{1/2}$  &  35.3  &  35.4   &  35.4  &   35.5  \\
$^{20}$B$(s_{1/2}, 2)$  & $s_{1/2}$  &  58.8  &  58.9   &  59.1  &   59.2  \\
\hline
\end{tabular}
  \caption{Weight ($W$) in \% of the components included in the
    calculation of $\frac{3}{2}^-$ bound state of $^{21}$B for each of
    the I-IV potentials.  The upper part corresponds to the first
    Jacobi set, where $\ell_x$ is the relative momentum between the
    two neutrons, $j_x$ is the dineutron total angular momentum, and
    $\ell_{j_y}$ is the relative wave between the core and the
    dineutron (keep in mind that the core has spin $3/2$). The lower
    part corresponds to the second and the third Jacobi sets, where
    $\ell_{j_n}$ is the shell occupied by the neutron in $^{20}$B,
    $j_x$ is the total angular momentum of $^{20}$B, and $\ell_{j_y}$
    is the relative wave between the second neutron and $^{20}$B.
    Only components contributing more than 2\% are shown on the
    table.}
\label{tableres1}
\end{table}

\subsection{$\frac{3}{2}^-$ bound state}

An analysis of the contribution to the norm of the wave function of
the different partial waves for the bound states in the first row of
Table~\ref{tableen} is given in Table~\ref{tableres1}. The upper part
of the table refers to the components in the first Jacobi set, where
$\ell_x$ is the relative momentum between the two neutrons, $j_x$ is
the dineutron total angular momentum, and $\ell_{j_y}$ is the relative
wave between the core and the dineutron (keep in mind that the core
has spin $3/2$). We can see that basically the full wave function
corresponds to the two neutrons in a relative $s$-wave ($\ell_x$=0),
and the $^{19}$B-core also in an $s$-wave relative to the center of
mass of the two neutrons ($\ell_y$=0).

The same kind of analysis is made in the lower part of the table in
the second and third Jacobi sets. Here $\ell_{j_n}$ indicates the
shell occupied by the neutron in $^{20}$B, $j_x$ is the total angular
momentum of $^{20}$B, and $\ell_{j_y}$ is the relative wave between
the second neutron and $^{20}$B.  Since the core is about 19 times
heavier than the neutron, the three-body center of mass is very close
to the core center of mass, and $j_{\ell_y}$ is therefore very close
to the shell occupied by the second neutron relative the
$^{19}$B-core. As we can see, the four potentials give rise to a very
similar structure, with about 95\% of the wave function corresponding
to both neutrons in the $s_{1/2}$ shell.

Note that, since $\ell_x$=$\ell_y=$0, the component shown in the upper
part of Table~\ref{tableres1}, which gives almost 100\% of the wave
function, can also be written as
$|(\frac{1}{2},\frac{1}{2})j_x=0,s_c=\frac{3}{2}, J=\frac{3}{2}
\rangle$. This component can be easily rotated into the second (or
third) Jacobi set as:
\begin{eqnarray}
\lefteqn{ \hspace*{-1cm}
\Big|\left(\frac{1}{2},\frac{1}{2}\right)j_x=0,s_c=\frac{3}{2}, J=\frac{3}{2} \Big\rangle
= } \nonumber \\ & &
-\sqrt{\frac{3}{8}} \Big|\left(\frac{3}{2},\frac{1}{2}\right)j_x=1,s_n=\frac{1}{2}, J=\frac{3}{2} \Big\rangle \\ & &
+\sqrt{\frac{5}{8}} \Big|\left(\frac{3}{2},\frac{1}{2}\right)j_x=2,s_n=\frac{1}{2}, J=\frac{3}{2} \Big\rangle \nonumber,
\end{eqnarray}
whose coefficients (squared) agree pretty well with values quoted in
the lower part of the table.

At this stage, it is important to keep in mind the uncertainty, always
present in this kind of three-body calculations, arising from those
aspects of the interaction going beyond the bare two-body potentials,
and which, as already mentioned, are typically accounted for by means
of a three-body potential ($V_\mathrm{3b}(\rho)$ in
Eq.(\ref{couprad})). The choice of this potential is to some extent
arbitrary, and in this work it has been taken such that the lowest
computed resonance is located at the experimental value. Of course,
this choice can be changed, and we could instead use a three-body
potential such that the lowest computed states, the bound states, are
pushed up to the experimental resonance energy. In order to do so, the
only thing to be done is to reduce the attraction in the three-body
potential, making it repulsive if necessary, such that the bound state
crosses the three-body threshold becoming a resonance. With the
appropriate choice for $S_\mathrm{3b}$ in Eq.(\ref{3bpot}), it should
in principle be possible to locate the lowest $\frac{3}{2}^-$ state at
the experimental value.

\begin{figure}
   \begin{center}
\includegraphics[width = 0.45\textwidth]{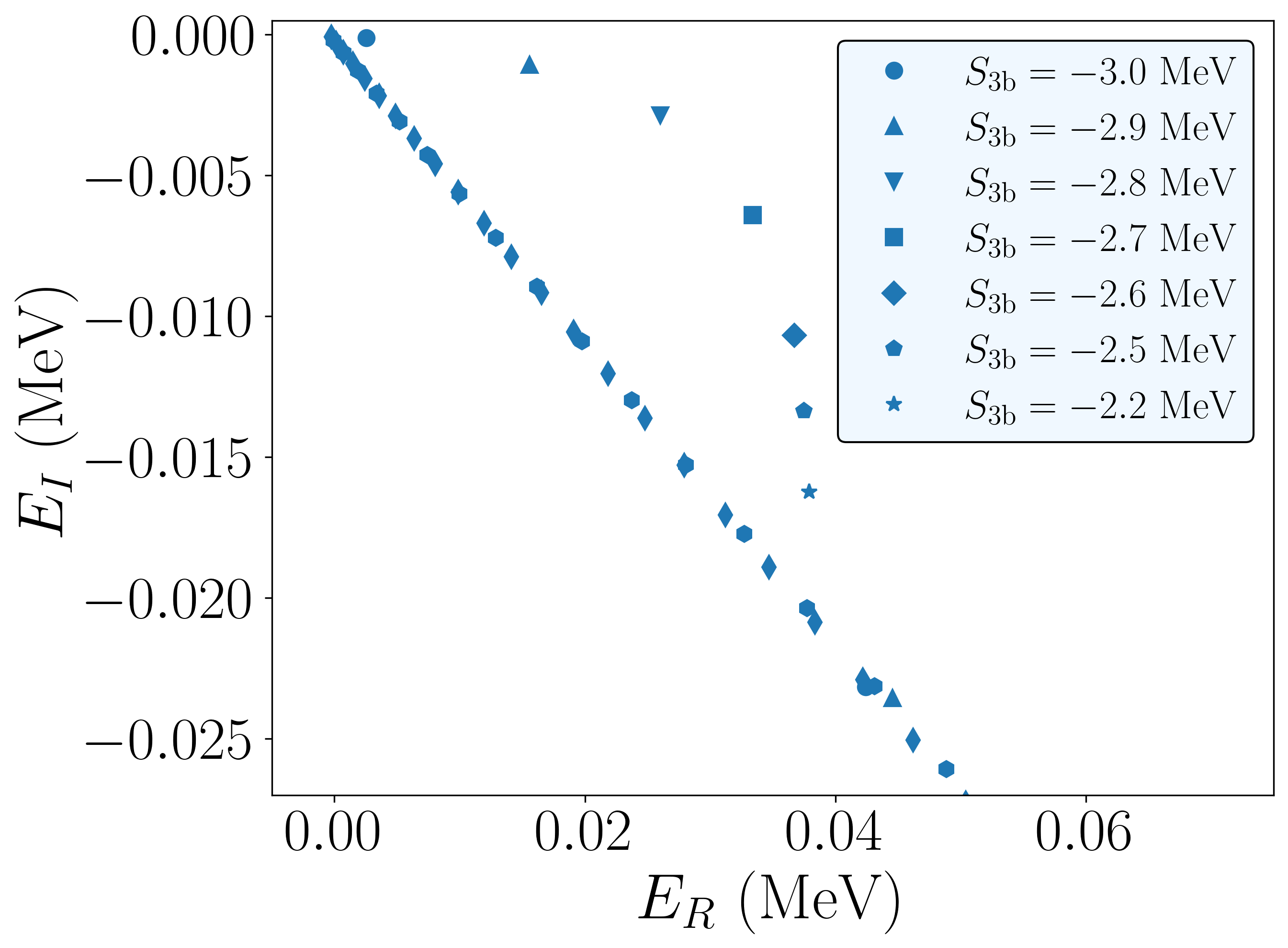}
   \end{center}
   \vspace*{-0.5cm}
   \caption{The real ($E_R$) and imaginary ($E_I$) parts of the energy
     for $\frac{3}{2}^-$ states in $^{21}$B (potential I) obtained
     after complex rotation for several different strengths,
     $S_\mathrm{3b}$, of the three-body potential. }
   \label{bound}
\end{figure}

When this is done, we obtain what shown in Fig.~\ref{bound} for
potential I (a similar result is obtained for the other three
potentials). When the attraction in the three-body potential is
reduced from $S_\mathrm{3b}$=$-11.9$ MeV to $-3.0$ MeV, the
$\frac{3}{2}^-$ bound state moves up in energy from the bound state at
$-3.33$ MeV to a resonance just above the three-body threshold shown
by the circle in Fig.~\ref{bound}. Starting from this point, and
making the three-body force less and less attractive, we can then
follow how this resonance moves in the energy plane. What we find is
that the resonance width increases very fast, and for the three-body
strength $S_\mathrm{3b}$=$-2.2$ MeV (star in the figure) the resonance
is about to disappear in the three-body continuum. Even more, contrary
to the width, the energy of the resonance increases very slowly, and
it is not possible to make it larger than about 0.05 MeV, very far
from the experimental value of 2.47 MeV.

The conclusion here is therefore that a $\frac{3}{2}^-$ state in
$^{21}$B below the experimentally known resonance at 2.47 MeV, and
coming from the two valence neutrons in the $s_{1/2}$-shell, should
actually exist. However, due to the uncertainty arising from the
three-body potential, this state might not necessarily be
bound. Furthermore, if the state exists as a resonance, either the
energy is extremely small close the three-body threshold, or the
resonance becomes soon very wide and probably out of experimental
reach for resonance energies larger than about 50 keV.

Since the bound $\frac{3}{2}^-$ state in $^{21}$B has not been
observed so far, we can consider those potentials giving rise to a
less bound state as the most appropriate ones for an accurate
description of $^{21}$B. As we can see in Table~\ref{tableen}, this is
what happens with potentials I and II, which are as well the ones
producing a resonance width closer to the experimental value.

\subsection{$\frac{3}{2}^-$ resonances}

Since, as previously shown, the computed bound state in $^{21}$B can
not be pushed up to the energy of the experimentally known resonance,
we can then conclude that this resonance is actually described by the
second state in Table~\ref{tableen}. The energy of this state has been
adjusted to the experimental value by means of the three-body
potential in Eq.~(\ref{3bpot}).

\begin{table}
\begin{tabular}{cc|cccc}

$^{20}$B($\ell_{j_n}$, $j_x$) & $\ell_{j_y}$ &$W^\mathrm{I}$ &  $W^\mathrm{II}$  & $W^\mathrm{III}$  &  $W^\mathrm{IV}$\\ \hline \hline
$^{20}$B$(s_{1/2}, 1)$  & $d_{3/2}$  &  27.3  &  17.8   &  43.1  &  36.9  \\
$^{20}$B$(s_{1/2}, 2)$  & $d_{3/2}$  &  18.6  &  27.0   &  0.0  &   0.0  \\
$^{20}$B$(d_{3/2}, 1)$  & $s_{1/2}$  &  42.2  &  45.1   &  15.4  &   20.0  \\
$^{20}$B$(d_{3/2}, 2)$  & $s_{1/2}$  &  5.2  &  1.1   &  28.9  &   18.8  \\
\hline
\end{tabular}
  \caption{In the second and third Jacobi sets, the same as
    Table~\ref{tableres1} for the $3/2^-$ resonance at 2.47 MeV. Only
    the components contributing with at least 3\% in at least one of
    the potentials are shown.}
\label{tableres2}
\end{table}

When written in the first Jacobi set, the resonance wave function
contains non-negligible contributions for quite a few different
components. Among these components the one with the dineutron in a
relative $s$-wave and the core in a relative $d_{3/2}$-wave
dominate. However many other contribute, and the first Jacobi set is
then not appropriate to clearly disentangle the structure of the
resonance.

The situation is different when the wave function is written in the
second or third Jacobi sets. In this case only four components give
more than 80\% of the wave function. These components, and their
weight for each of the four potentials used in this work, are given in
Table~\ref{tableres1}. All the other components not shown in the table
never contribute with more than 3\% to the resonance wave function. In
the table we can see that in all the cases the resonance is formed by
one neutron in the $s_{1/2}$-shell and the second one in the
$d_{3/2}$-shell.

Entering into the detail, we observe that for potentials I, II, and
III, the weights of the components agree with the $^{20}$B spectrum
shown in Fig.~\ref{niveles} produced by each of the potentials. For
potentials I and III the component with $^{20}$B in an $s_{1/2}$-wave
and $j_x$=1 dominates over the component with $j_x$=2, and the other
way around for potential II. The same happens for the component with
$^{20}$B in a relative $d_{3/2}$-wave. For potentials I and II the
component with $j_x$=1 dominates over the case with $j_x$=2, and the
other way around for potential III. Note that $j_x$=0 and $j_x$=3 are
not possible, since they can not produce the total angular momentum
$3/2$ when coupling to an $s_{1/2}$-neutron. Therefore, the lowest of
the active $^{20}$B resonances coming from the $d_{3/2}$-wave is
$j_x$=1 for potentials I and II and $j_x$=2 for potentials III and IV.

However, as we can see in Table~\ref{tableres1}, this rule does not
apply for potential IV. According to the level scheme in
Fig.~\ref{niveles}, for this potential the components with the
$s_{1/2}$-neutron and the $d_{3/2}$-neutron coupled to $j_x=2$ should
dominate. What seen in the table is just the opposite, being the
components with $j_x=1$ the ones dominating. The explanation for this
can be obtained after analyzing the partial wave content of the second
$3/2^-$ resonance also visible in Fig.~\ref{spectr} for the four
potentials. The resonance energy is about 4.5 MeV for potentials I and
II, and in the vicinity of 3.3 MeV for potentials III and IV. The
structure of this resonance, which is shown in
Table~\ref{tablepartial}, is similar to the one of the first
resonance, with one neutron in the $s_{1/2}$-shell and the second
neutron in the $d_{3/2}$-shell. However, we immediately see that, for
potential IV, this second resonance is dominated by the $j_x=2$
components, which was the expected result for the first
one. Therefore, in the case of potential IV the interplay between the
neutron-core and the neutron-neutron interactions leads to a crossing
between the two resonances compared to the result with potentials I,
II, and III.

\begin{table}
\begin{tabular}{cc|cccc}
$^{20}$B($\ell_{j_n}$, $j_x$) & $\ell_{j_y}$ &$W^\mathrm{I}$ &  $W^\mathrm{II}$  & $W^\mathrm{III}$  &  $W^\mathrm{IV}$\\ \hline \hline
$^{20}$B$(s_{1/2}, 1)$  & $d_{3/2}$  &   26.1  &  35.9   &  0.0  &  1.9  \\
$^{20}$B$(s_{1/2}, 2)$  & $d_{3/2}$  &   23.2  &  19.5   &  45.9 &  46.2  \\
$^{20}$B$(d_{3/2}, 1)$  & $s_{1/2}$  &    2.6  &   0.2   &  26.6 &  20.0  \\
$^{20}$B$(d_{3/2}, 2)$  & $s_{1/2}$  &   47.1  &  57.0   &  19.9 &  28.4  \\
\hline
\end{tabular}
  \caption{The same as Table~\ref{tableres2} for the second $3/2^-$
    resonance. Only the components contributing with at least 3\% in
    at least one of the potentials are shown.}
\label{tablepartial}
\end{table}

The computed overall structure of the $3/2^-$ resonances was to some
extent an expected result. Since we have found a $^{21}$B ground state
built by locating the two neutrons in the $s_{1/2}$-wave, it is
reasonable to think that the first excited state will be produced by
the jump of one of those neutrons into the next available shell, i.e.,
the $d_{3/2}$-shell. The different possible couplings between the spin
of the neutrons and the core make possible the existence of more than
one resonance with similar $sd$-structure.

\section{Energy distributions}

We compute the energy distributions for the three-body decay of
$^{21}$B by means of a Monte Carlo simulation. A large number of
random events ($5\times 10^5$) is generated, each consisting of three
four-momenta corresponding to the three decaying fragments. Each event
must satisfy the condition that the sum of the center-of-mass energies
of the fragments equals the resonance energy $E_r=2.47$~MeV. The
probability of each event is set to be the absolute-squared wave
function at a large distance, chosen to be $\rho = 90$~fm, where the
asymptotic behavior has been already reached.  We give the energies in
units of their maximum values for each case (i.e.,
$E_\mathrm{c}^\mathrm{max}$=2/21$E_r$ for the core and
$E_\mathrm{n}^\mathrm{max}$=20/21$E_r$ for the neutrons).  Once all
this information is obtained, different kinds of plots can be
produced, providing different insights into the decay process.

\begin{figure}
  \begin{center}
    \includegraphics[width = 0.45\textwidth]{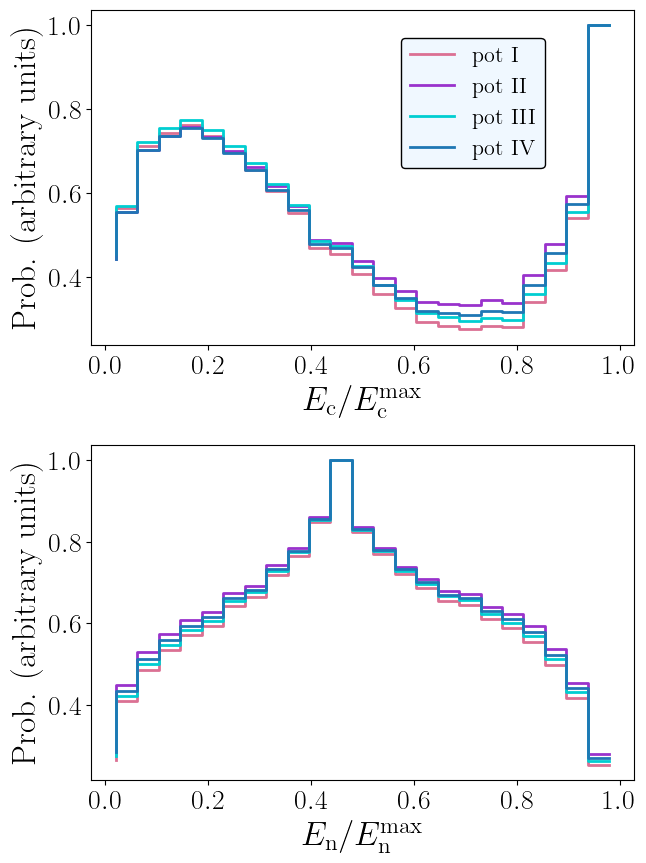}
  \end{center}
  \caption{ Energy distributions of the $^{19}$B core ($c$, top panel)
    and the neutron ($n$, bottom panel) for the $\frac{3}{2}^-$
    resonance at 2.47~MeV in $^{21}$B, calculated using the four
    interaction potentials under consideration (pot I-IV). In both
    panels, the energies are normalized to their maximum allowed
    values. }
  \label{Ecoren4pots}
\end{figure}

In the upper panel of Fig.~\ref{Ecoren4pots} we present the
probability of a given core energy,
$E_\mathrm{c}/E_\mathrm{c}^\mathrm{max}$. A broad peak is observed at
$E_\mathrm{c}\approx 0.2 E_\mathrm{c}^\mathrm{max}$. This peak is
consistent with a process where most of the energy is taken by the
relative neutron-neutron energy, meaning therefore a direct three-body
decay of the resonance. Beyond this peak, the distribution decreases
to a minimum around $E_\mathrm{c}\approx 0.75
E_\mathrm{c}^\mathrm{max}$, followed by a narrow enhancement when the
$E_\mathrm{c}$ approaches its maximum value. This high energy peak
represents a process where the core and the two-neutrons are emitted
in opposite directions, i.e., a two-neutron decay process.

Therefore, the resonance decay mechanism is a combination of direct
decay and two-neutron emission, with no traces of sequential decay
through any $^{20}$B intermediate state. This is more easily seen in
the lower part of Fig.~\ref{Ecoren4pots}, where we display the neutron
energy distribution. This distribution starts at zero and returns
again to zero at the maximum energy, being nearly symmetric and
reaching its maximum at approximately half of the maximum energy. This
means that the neutrons have a preference to equally share the
available energy. This is an indication that a sequential decay
mechanism is not present, since in that case the neutron energy
distribution should present a prominent peak (typically at low energy)
matching with the energy of the core-neutron resonance, plus an
additional peak (typically at high energy) corresponding to the second
neutron taking the remaining available neutron energy. Our finding is
in agreement with Ref.\cite{leb18}, where $^{21}$B is considered a
potential new case of direct two-neutron decay.

In the figure, the four potentials considered are represented by
different colors; however, their corresponding distributions are
nearly indistinguishable.

In a three-body decay, the available energy is shared among the three
decaying fragments, giving rise to continuous single-particle energy
distributions, as in Fig.~\ref{Ecoren4pots}. These distributions alone
do not provide a complete description of the decay kinematics. Since
the total resonance energy is conserved, a full characterization
requires the simultaneous knowledge of the energies of two of the
emitted particles. This information is conveniently represented
through two-dimensional energy correlation diagrams, commonly known as
Dalitz plots, which were introduced initially to study decays of
$K$-mesons \cite{Dal53}. These plots provide direct insight into the
correlations among the decay fragments and have become a standard tool
for investigating the dynamics of three-body decays. We use the same
Monte Carlo technique as for the individual particle energy
distributions.

\begin{figure}
  \begin{center}
    \includegraphics[width = 0.4\textwidth]{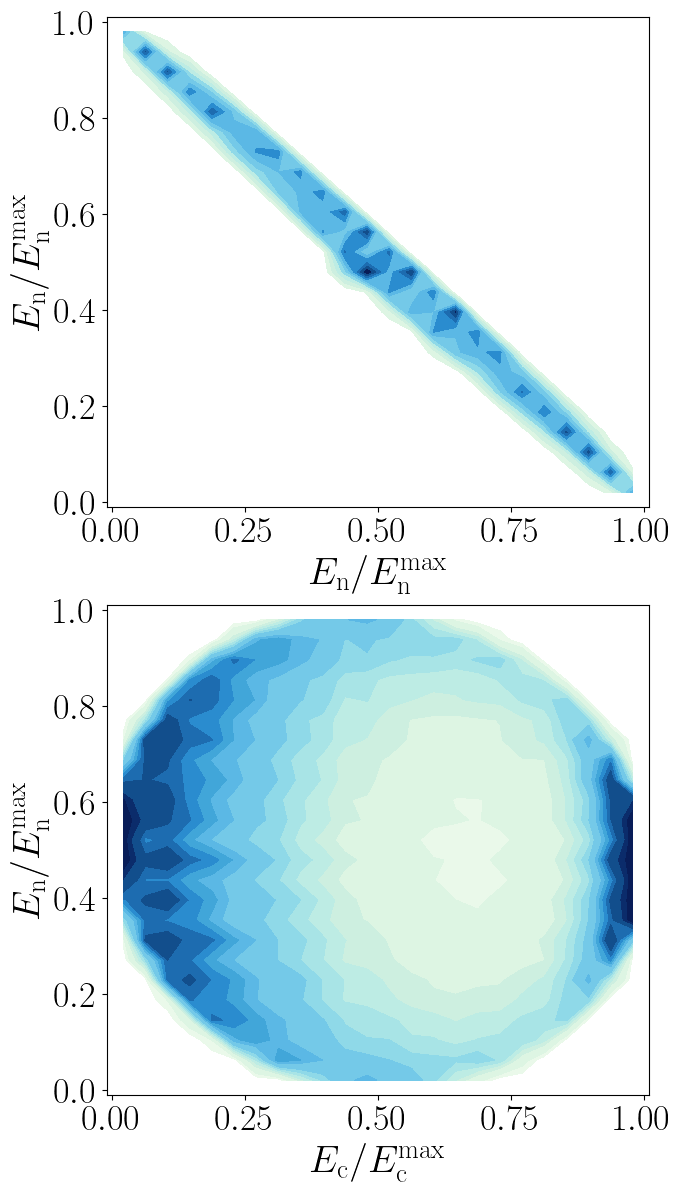}
  \end{center}
  \caption{Dalitz plots of the energy-correlation distributions for
    the $\frac{3}{2}^-$ resonance of $^{21}$B at 2.47~MeV obtained
    with potential considered I. The upper plot shows the energy of
    one neutron versus the energy of the other neutron. The lower plot
    shows instead the energy of the core on the $x$-axis versus the
    energy of one neutron on the $y$-axis. The results have been
    calculated using $5\times10^5$ Monte Carlo events. }
  \label{2D_potI}
\end{figure}

Fig.~\ref{2D_potI} shows the two-dimensional energy correlations (or
Dalitz plots) for the three-body decay of the lowest $\frac{3}{2}^-$
resonance of $^{21}$B. The upper panel displays the neutron-neutron
($n$-$n$) correlations, while the lower panel corresponds to the
$^{19}$B-neutron ({\it core}-$n$) correlations. As in the previous
figure, the energies are normalized to their respective maximum
kinematically allowed values. Only the results obtained with potential
I are shown, since the four interaction potentials considered lead to
virtually identical correlation patterns. The distributions provide a
basis for direct comparison with future experimental data.

As expected, the $n$-$n$ and core-$n$ correlation diagrams are
symmetric under the exchange of the two neutrons, since they are
indistinguishable. No forbidden regions are observed in the allowed
kinematic space, but the probability reaches a minimum when the core
carries approximately $0.7$ of its maximum energy and one neutron
about $0.5$ of its maximum energy. Since angular momentum conservation
does not exclude any kinematically allowed energy combination, the
entire Dalitz plot is populated. The probability density is enhanced
when the core carries either a small or a large fraction of the
available energy and the neutron about half of it.

\section{Search for other resonances}

Despite the scarce experimental information available for $^{21}$B, we
have explored the possible existence of additional states, in
particular the $1/2^-$ states. In principle positive parity states
could also be possible. However, that would require a reasonably
accurate description of the $p$-wave potential between $^{19}$B and
the neutron, which, due to the absence of experimental information
about positive parity states in $^{20}$B, in this work has been put
equal to zero.

\begin{figure}
  \begin{center}
    \includegraphics[width = 0.4\textwidth]{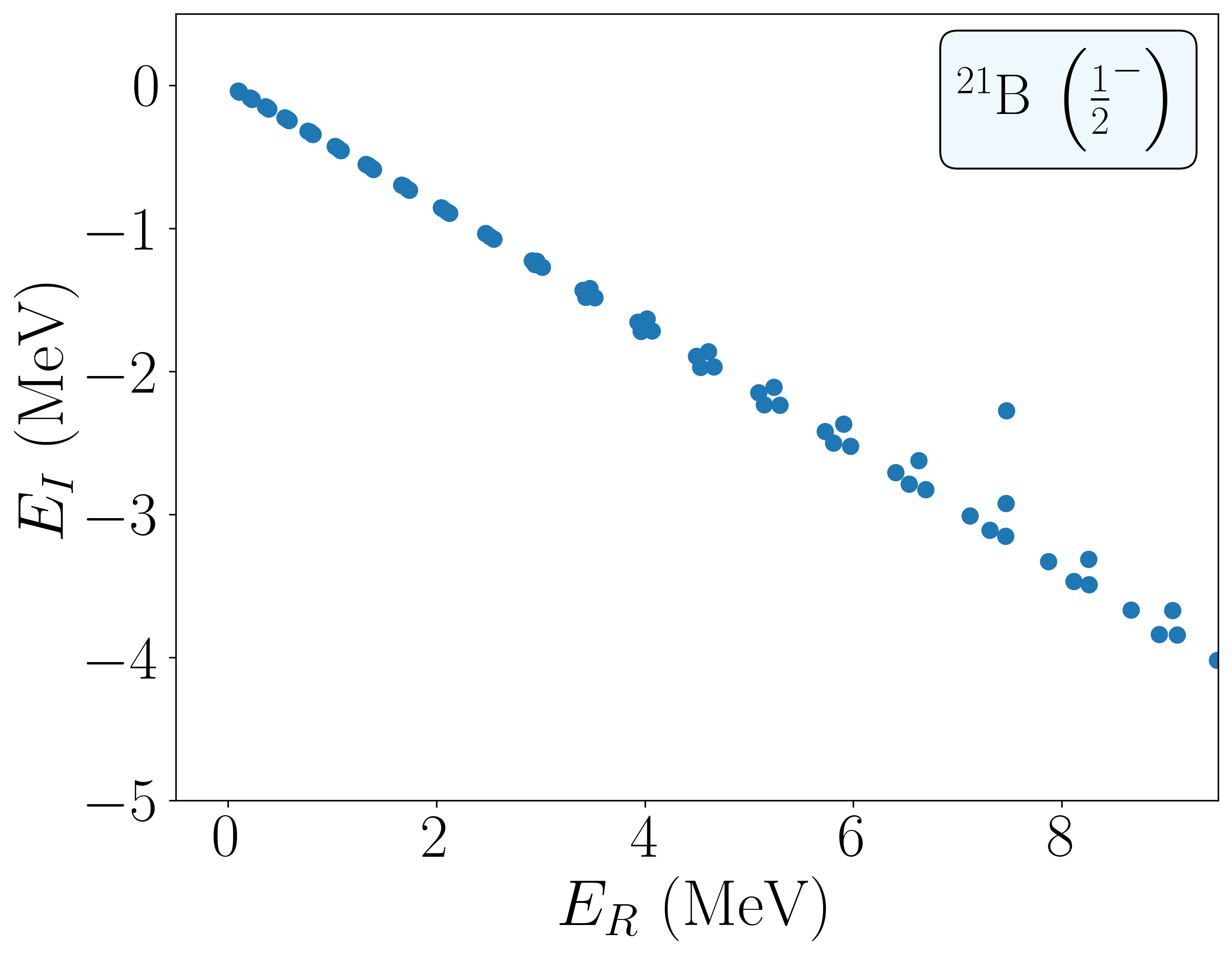}
  \end{center}
  \caption{ The real ($E_R$) and imaginary ($E_I$) parts of the
    energies for the $1/2^-$ state in $^{21}$B obtained after complex
    rotation with angle $\theta = 0.2$~rad. Potential I has been
    used.}
  \label{other}
\end{figure}

As done in Fig.~\ref{spectr} for the $3/2^-$ states, we show in
Fig.~\ref{other} the result obtained for the 1/2$^-$ states using
potential I and a complex-scaling angle $\theta=0.2$ rad. As we can
see, a resonant state shows up with energy and width of about 7.5 MeV
and 4.5 MeV, respectively. It is important to keep in mind that this
result has been obtained without use of any three-body potential,
since the absence of experimental constraints prevents any meaningful
tuning.

The computed $1/2^-$ resonance contains a large contribution (about
40\%) of $^{20}$B in the high-lying $0^-$ resonance, with the second
neutron in a relative $s_{1/2}$-wave. In any case, the current
experimental information is insufficient to truly establish the
existence of this state, and our results suggest that further
experimental investigations of $^{21}$B would be highly valuable.

\section{Summary and conclusions}

In this work, the structure of the neutron-rich unbound nucleus
$^{21}$B has been investigated within a three-body model consisting of
a $^{19}$B core and two neutrons. The three-body wave functions have
been obtained using the hyperspherical adiabatic expansion method
combined with the complex-scaling technique, allowing a consistent
treatment of both bound states and resonances. Four different
potential sets compatible with the known experimental spectrum of
$^{20}$B have been considered. The $s$-wave potential has been chosen
shallow enough such that the neutron can not be bound in the Pauli
forbidden state. For $d$-waves a large and positive spin-orbit
strength has been used to push the $d_{5/2}$-shell out the active
space. For $p$-waves the interaction has been put equal to zero.

The three-body calculation reveals that at least three different
3/2$^-$ states should appear in the $^{21}$B low-lying spectrum. The
ground state is consistent with the two neutrons occupying the
$s_{1/2}$-shell, whereas the two excited states correspond to one
neutron in the $s_{1/2}$-shell and the second one in the
$d_{3/2}$-shell. Although not found in this work, in principle higher
resonant states with the two neutrons in the $d_{3/2}$-shell could
also be present.

We have found that only when a three-body potential is used to locate
the lowest computed resonance at the experimental energy, the ground
state, which is actually bound, shows up. In fact, the analysis of the
three-body force dependence shows that this state cannot be
continuously shifted to the experimentally observed resonance
energy. As the attraction of the three-body force is reduced, the
state becomes a near-threshold resonance whose width increases rapidly
while its energy remains very close to the three-body
threshold. Consequently, our calculations indicate that a $3/2^-$
state, with the two neutrons in the $s_{1/2}$-shell, should exist
below the observed resonance. Although uncertainties in the three-body
interaction do not allow us to determine whether this state is
strictly bound or an extremely low-energy resonance, the present
results strongly suggest that it is distinct from the experimentally
known state at 2.47 MeV.

The experimentally observed resonance is interpreted as the second
$3/2^{-}$ state of the system. Its calculated width is of the order of
0.75-1~MeV, in reasonable agreement with the experimental data. It
corresponds with and $s_{1/2}$-$d_{3/2}$ configuration. The predicted
momentum-correlation distributions are largely independent of the
choice of the neutron-core potential, indicating that the decay
observables are more determined by the three-body dynamics than by
details of the two-body interaction. In particular, the calculations
predict a resonance decay mechanism combining a direct three-body
decay and two-neutron emission. No traces of sequential decay through
any $^{20}$B state are found.

Finally, a search for additional states predicts the presence of
$1/2^-$ resonance with an excitation energy of about 7.5 MeV. No bound
states were found for these spin-parity assignment. The current
experimental information is insufficient to confirm their existence,
but future measurements could provide valuable tests of these
predictions

\section{Acknowledgments}

This work has been supported by Grants No. PID2022-136992NB-I00 and
PID2021-122711NB-C21 funded by MCIN/AEI/10.13039/501100011033, Spain.

\end{document}